\documentclass[runningheads,citeauthoryear]{apinv}
\usepackage{epsfig,cite,graphics}

\begin{document}

\title{Magnetic Field and Activity of the Single Late-type Giant $\beta$ Ceti\thanks{Based on observations obtained at the Bernard Lyot T\'elescope (TBL, Pic du Midi, France) of the Midi-Pyr\'en\'ees Observatory which is operated by the Institut National des Sciences de l'Univers of the Centre National de la Recherche Scientifique of France, and at the Canada-France-Hawaii Telescope (CFHT) which is operated by the National Research Council of Canada, the Institut National des Sciences de l'Univers of the Centre National de la Recherche Scientifique of France, and the University of Hawaii.}}
\titlerunning{Magnetic field and Activity of $\beta$ Ceti}
\author{S. Tsvetkova\inst{1}, M. Auri\`ere\inst{2}, R. Konstantinova-Antova\inst{1,2}, G.A. Wade\inst{3}, R.G. Bogdanovski\inst{1}, P. Petit\inst{2}}
\authorrunning{Tsvetkova et al.}
\institute{Institute of Astronomy and NAO, Bulgarian Academy of Sciences, 72 Tsarigradsko shose, 1784 Sofia, Bulgaria\\
              \email{stsvetkova@astro.bas.bg}
	\and CNRS, Institut de Recherche en Astrophysique et Plan\'etologie, 14 Avenue Edouard Belin, 31400 Toulouse, France\\
	\and Department of Physics, Royal Military College of Canada, PO Box 17000, Station `Forces', Kingston, Ontario, Canada K7K 4B4\\}
\papertype{Conference talk}
\maketitle

\begin{abstract}
We present the behavior of the magnetic field and activity indicators of the single late-type giant $\beta$ Ceti in the period June 19, 2010 -- December 14, 2010. We used spectropolarimetric data obtained with two telescopes --- the NARVAL spectropolarimeter at T\'elescope Bernard Lyot, Pic du Midi, France and the ESPaDOnS spectropolarimeter at CFHT, Hawaii. The data were processed using the method of Least Square Deconvolution which enables to derive the mean photospheric profiles of Stokes I and V parameters. We measured the surface-averaged longitudinal magnetic field B$_{l}$, which varies in the interval 0.1 - 8.2 G, the line activity indicators CaII K, H$\alpha$, CaII IR (854.2 nm) and radial velocity. By analyzing the B$_{l}$ variations, was identified a possible rotational period P = 118 days. A single, large magnetic spot which dominates the field topology is a possible explanation for the B$_{l}$ and activity indicator variations of $\beta$ Ceti.
\end{abstract}


\section{Introduction}
$\beta$ Ceti (HD 4128, HR 188, HIP 3419) is a single late-type giant star of spectral class K0 III with $V = 2.04$ mag and $B-V = 1.02$. It has started as an A-type star on the main sequence (Maggio et al., 1998). The fundamental parameters of $\beta$ Ceti and their references are presented in Table\ref{table1}.

\begin{figure}
  \begin{center}
    \centering{\epsfig{file=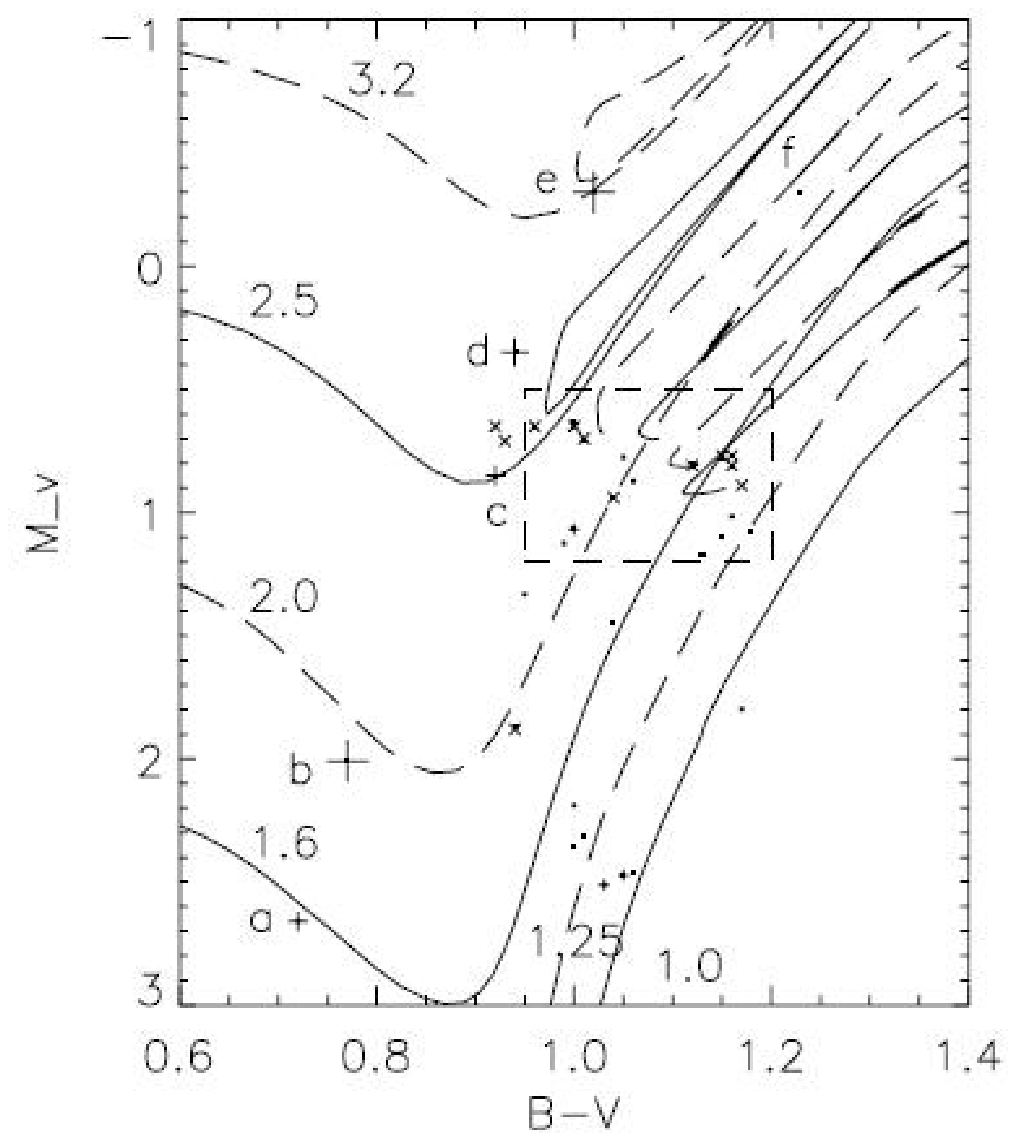, width=0.4\textwidth}}
    \caption[]{HR diagram by Schr\"{o}der et al. (1998). Plus symbols are RASS detections. $\beta$ Ceti is pointed out as ``+'' and letter ``e'' with mass of 3.2 $M_{\sun}$. The giant clump region is indicated by the rectangle.}
    \label{figure1}
  \end{center}
\end{figure}

There is discussion in the literature about its evolutionary status: is it a first-ascent red giant or a clump giant star? According to Schr\"{o}der et al. (1998) clump giants are not visible in the RASS (ROSAT All-Sky Survey). In this study they use data for $\beta$ Ceti from RASS detections. On their HR diagram they pointed out the giant ``clump region'' (Fig.\ref{figure1}). It is obvious from this figure that the evolutionary track of $\beta$ Ceti is far away from this region and the star is rather possibly situated at the base of the red giant branch.

According to other authors $\beta$ Ceti is a He-burning clump giant because of its photospheric abundances. Its low Li abundance $\log \epsilon (Li) = 0.01$ and low values of the ratios $C/N = 1.38$ and $C^{12}/C^{13} = 19 \pm 2$ (Luck \& Challener, 1995; Tomkin et al., 1975) suggest this classification. The FUV spectral properties of $\beta$ Ceti are very similar to other active clump stars like the Hyades giants $\theta^{1}$ Tau and $\gamma$ Tau, and the Capella G8 III primary (Ayres et al., 1983; Ayres et al., 1998). Another survey of photospheric abundances of $\beta$ Ceti was carried out by S\"{a}gesser \& Jordan (2005). They compared the results from different authors about the abundances of Fe, N, O, Mg and Si. According to these results, their suggestion is that this star has already undergone first dredge-up.

\begin{table}
\begin{center}
\caption{The fundamental parameters of $\beta$ Ceti and their references}
\begin{tabular*}{0.95\textwidth}{@{\extracolsep{\fill}} p{0.30\textwidth} p{0.22\textwidth} p{0.37\textwidth} }
\hline
Parameter	&	Interval of values	&	References\\
\hline                                                                    

Mass       &  2.8 -- 3.2 $M_{\sun}$  &  Maggio et al., 1998; Schr\"{o}der et al., 1998; Gondoin, 1999; Allende Prieto \& Lambert, 1999; Berio et al., 2011\\
Radius     &  $\sim$ 15 $R_{\sun}$  &  Jordan \& Montesinos, 1991; Allende Prieto \& Lambert., 1999; S\"{a}gesser \& Jordan, 2005\\
Parallax and Distance   &  $33.86\pm0.16$ mas $29.38 ^{+0.63} _{-0.69}$ pc & from Hipparcos Catalogues - Perryman et al., 1997; van Leeuwen, 2007\\
Effective Temperature &   4750 -- 5000 K       &  Jenkins, 1952; Ayres et al., 1998; Gondoin, 1999; Luck et al., 1995; Eriksson et al., 1983; Ottmann et al., 1998; Gratton \& Ortolani, 1986\\
$\log g$    &  2.45 -- 3.05         &  Jenkins, 1952; Jordan \& Montesinos, 1991; Eriksson et al., 1983; Luck et al., 1995; Ottmann et al., 1998; Gratton \& Ortolani, 1986\\
$[Fe/H]$    &  -0.09 -- 0.13      &  Luck et al., 1995; Gondoin, 1999; Alves, 2000; Ottmann et al., 1998; Gratton \& Ortolani, 1986\\
$v \sin i$  &   2.5 -- 4 km/s       &  Fekel, 1997; Carney et al., 2008; Smith \& Dominy, 1979; Gray, 1982; Ottmann et al., 1998\\
Inclination angle i   &	  $60^\circ$   &  Sanz-Forcada et al., 2002, Gray, 1989\\
Rotational period &  199 days    &  Jordan \& Montesinos, 1991\\
                  &  171 days    &  C. Jordan, private comunication\\
\hline
\end{tabular*}
\label{table1}
\end{center}
\end{table}

$\beta$ Ceti is classified as a single giant star with the highest X-ray luminosity $\log L_{x} = 30.2$ erg/s in the solar neighborhood ($d \le 30$ pc), calculated with a distance of $d = 29.4$ pc (Maggio et al., 1998; H\"{u}nsch et al., 1996). With this high activity, the star is compared to Capella ($\alpha$ Aur), a long period RS CVn-type active binary, and $\theta^{1}$ Tau (K0 III). A model of the atmosphere of $\beta$ Ceti was performed by Eriksson et al. (1983) and they suggested the existence of coronal loops. Ayres et al. (2001) reported about a series of striking coronal flare events observed with EUVE during a period of 34 days starting on 1 Aug 2000.

$\beta$ Ceti has $V-R = 0.72$ which places it to the left of the coronal dividing line in the HRD. This nearly vertical dividing line was proposed by Linsky \& Haisch (1979) near $V-R = 0.80$ and it separates the stars into two groups --- the solar-type group ($V-R < 0.80$) which indicates the existence of a chromosphere, a transition region and a corona, while the non-solar-type group ($V-R > 0.80$) shows only chromosphere lines. The existence of the dividing line is confirmed by the study of Simon et al. (1982) with a larger sample of stars.

Using spectropolarimetric data, one can detect the presence of a magnetic field in plasma by the Zeeman effect (Zeeman, 1897) and measuring the Stokes parameters [I, Q, U, V]. For stars with weak magnetic fields it is possible to detect magnetic features only in Stokes V (circular polarization) because the signatures in Stokes Q and U (linear polarization) are much weaker than these in Stokes V. In our study for $\beta$ Ceti, we use Stokes I and V parameters which give information about the strength of the surface-averaged longitudinal magnetic field B$_{l}$.

\section{Observations and data reduction}
Observations were carried out at two telescopes --- the 2-m T\'elescope Bernard Lyot (TBL) at Pic du Midi Observatory, France and the 3.6-m Canada-France-Hawaii Telescope (CFHT) at Hawaii.

TBL and CFHT use the twin new-generation fiber-fed echelle spec\-tro\-po\-la\-ri\-me\-ters NARVAL (Auri\`ere, 2003) and ESPaDOnS (Donati et al., 2006) respectively, which in polarimetric mode have a spectral resolution of about 65 000 and a spectrum coverage from near-ultraviolet 370 nm to near-infrared 1050 nm in a single exposure. Stokes I (unpolarised light) and Stokes V (circular polarization) parameters were obtained by four sub-exposures between which the retarders -- Fresnel rhombs -- were rotated in order to exchange the beams in the instrument and to reduce spurious polarization signatures (Semel et al., 1993).

The raw data were processed using the automatic reduction software LibreEsprit, developed for NARVAL and ESPaDOnS. Additional details regarding the observing procedure and data reduction can be found in Donati et al. (1997).

For further processing of the data we used the Least Squares Deconvolution (LSD) technique (Donati et al., 1997). This technique enables averaging of several thousand absorption lines taken from one spectrum, which increases the signal-to-noise ratio (S/N) and thus allows us to detect weak magnetic signatures which would not be visible in individual lines. Using the LSD technique, one can derive mean photospheric profiles of Stokes I and Stokes V from a complete echelle spectrum.

\begin{figure}
  \begin{center}
    \centering{\epsfig{file=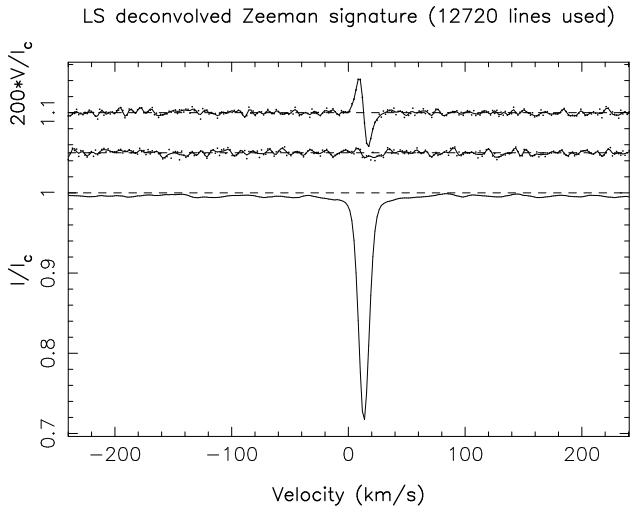, angle=-90, width=0.6\textwidth}}
\newline
\newline
\newline
    \caption[]{LSD profiles of $\beta$ Ceti observed on November 20, 2010 with ESPaDOnS. From top to bottom: mean Stokes V profile; diagnostic null spectrum; mean Stokes I profile. See details about these profiles in the text. Profiles are shifted vertically; the Stokes V profile and null spectrum are expanded by a factor of 200 for clarity.}
    \label{figure2}
  \end{center}
\end{figure}

Fig.\ref{figure2} shows the result after applying the LSD method to the spectrum of $\beta$ Ceti obtained on November 20, 2010. From top to bottom it depicts: the circularly polarized mean Stokes V profile; a diagnostic null spectrum; the unpolarized mean Stokes I profile. All profiles are normalized with respect to the continuum intensity. The diagnostic null spectrum serves to diagnose the presence of spurious contributions to the Stokes V spectrum.

The dataset includes observations in the period from June 19, 2010 to December 14, 2010 in which we have collected 25 spectra. First, spectra were automatically extracted using the LibreEsprit software and after that for the Zeeman analysis we used the LSD method with a mask calculated for an effective temperature $T_{eff} = 5000$ K, $\log g = 3.0$ and a microturbulence of 2 km/s, consistent with the literature data for the star (Thevenin, 1998; Hekker \& Mel\'endez, 2007). For each spectrum we calculated mean Stokes I and V profiles averaging about 12 700 lines. Then we computed the longitudinal magnetic field B$_{l}$ in G, using the first-order moment method (Donati et al., 1997; Rees \& Semel, 1979; Wade et al., 2000). We also measured the radial velocity (RV) on the LSD Stokes I profiles using a Gaussian fit. The RV stability of NARVAL and ESPaDOnS is about 30 m/s (Moutou et al., 2007). For better understanding of the stellar activity, we measured the activity indicators H$\alpha$ at 656.3 nm and the CaII IR triplet at 854.2 nm (relative intensities with respect to the continuum) and CaII K (the relative intensity of the emission core relative to the intensity at $\lambda = 355$ nm, I/I(395nm)).

In order to determine the rotational period of the star, we applied to our dataset a Lomb-Scargle period search analysis (Lomb, N.R., 1976; Scargle, J.D., 1982) for B$_{l}$ and some of the activity indicators.

\section{Results and Discussion}
The spectra allow us to obtain simultaneously the longitudinal magnetic field B$_{l}$, the activity indicators and the radial velocity so that we can study the correlations between them. The list of observations and measured values of B$_{l}$ in gauss, H$\alpha$, CaII K, CaII IR and their uncertainties, as well as the radial velocity measurements, are listed in Table\ref{table2}.

\begin{figure}
  \begin{center}
    \centering{\epsfig{file=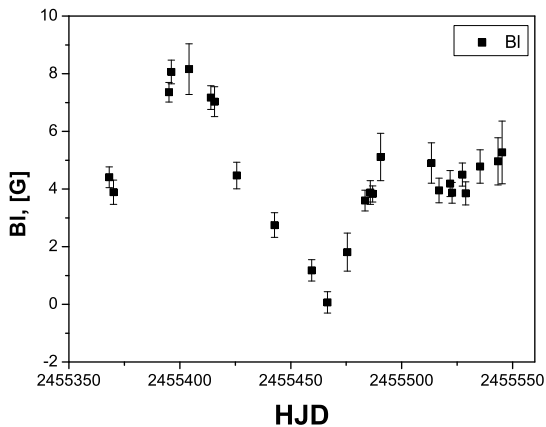, width=0.7\textwidth}}
    \caption[]{Time variations of the surface-averaged longitudinal magnetic field B$_{l}$ and its uncertainty in G.}
    \label{figure3}
  \end{center}
\end{figure}

The changes of the B$_{l}$ values are in the interval from 0.1 G to 8.2 G and B$_{l}$ remains of positive polarity for all our observations. The time variation of B$_{l}$ shows two types of behavior (Fig.\ref{figure3}). A sine wave is clearly visible in the first part of our observations in the period June 19, 2010 - October 20, 2010. We observe a plateau after the sinusoidal curve, which covers the period November 12, 2010 - December 14, 2010. The existence of this second part of the dataset, which is completely different from the first one, is an evidence for a change in the surface magnetic field configuration of the star.

\begin{figure}
  \begin{center}
    \centering{\epsfig{file=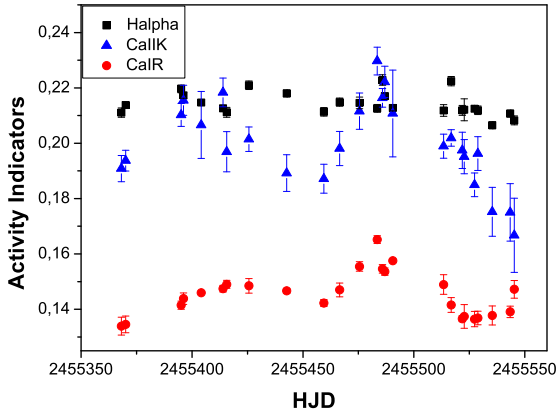, width=0.7\textwidth}}
    \caption[]{Time variations of the activity indicators H$\alpha$, CaII K and CaII IR from top to bottom, respectively, and their uncertainties.}
    \label{figure4}
  \end{center}
\end{figure}

The behavior of the activity indicators and their uncertainties are presented in Fig.\ref{figure4} and Table\ref{table2}. Their variations are in the intervals: 0.20 - 0.22 for H$\alpha$; 0.17 - 0.23 for CaII K; 0.13 - 0.16 for CaII IR. The activity indicators CaII K and CaII IR vary in nearly the same way but H$\alpha$ shows no significant variations with time, taking into account the accuracy of its measurements. Comparing the curves for B$_{l}$ and the activity indicators CaII K and CaII IR in Fig.\ref{figure3} and Fig.\ref{figure4} but only for the first part of the dataset, one can notice that the variability shows some similarity. This sinusoidal curve for B$_{l}$ could be due to a single large spot at the stellar surface but a more complex topology is also possible. Moreover CaII K and CaII IR also vary in a manner similar to B$_{l}$ which could support the idea of one big spot at the stellar surface which dominates the behavior of all magnetic activity indicators. It is not the same case for the second part of the dataset where H$\alpha$, CaII K, CaII IR show different behaviors regarding B$_{l}$. For this second part, the intensity of the indicators decrease which means that we detect less emission and the chromosphere is less heated. A smaller part of the hemisphere is occupied by the magnetic field.

\begin{figure}
  \begin{center}
    \centering{\epsfig{file=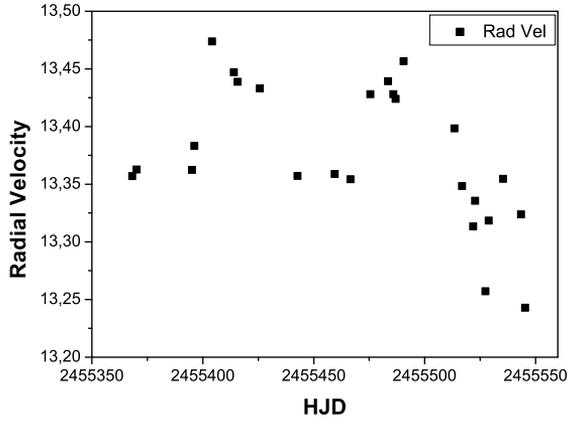, width=0.7\textwidth}}
    \caption[]{Time variations of the radial velocity. The stability of the instruments is about 30 m/s (Moutou et al., 2007).}
    \label{figure5}
  \end{center}
\end{figure}

Radial velocity variations with time are shown in Fig.\ref{figure5}. They vary in the interval 13.24 - 13.47 km/s.

\begin{table}
\begin{center}
\caption{Data for B$_{l}$, activity indicators and their accuracy}     
\label{table2}                                                       
\centering                                                            
\begin{tabular*}{0.95\textwidth}{@{\extracolsep{\fill}} c c c c c c c c c c}
\hline\hline                                                          
Date&  HJD& CaII K& $\sigma$& H$\alpha$& $\sigma$& CaII IR& $\sigma$& B$_{l}$, [G]& $\sigma$, [G]\\
\hline                                                                
19 Jun 10 & 2455368.123	& 0.191	& 0.005	& 0.211	& 0.002	& 0.134	& 0.003	& 4.4 & 0.4\\
21 Jun 10 & 2455370.134 & 0.194 & 0.004 & 0.214 & 0.001 & 0.135 & 0.003 & 3.9 & 0.4\\
16 Jul 10 & 2455395.140 & 0.210 & 0.004 & 0.220 & 0.001 & 0.141 & 0.001 & 7.4 & 0.3\\
17 Jul 10 & 2455396.132 & 0.216 & 0.005 & 0.217 & 0.001 & 0.144 & 0.002 & 8.1 & 0.4\\
25 Jul 10 & 2455404.138 & 0.207 & 0.012 & 0.215 & 0.001 & 0.146 & 0.000 & 8.2 & 0.9\\
04 Aug 10 & 2455414.006 & 0.218 & 0.005 & 0.213 & 0.001 & 0.147 & 0.001 & 7.2 & 0.4\\
06 Aug 10 & 2455415.633 & 0.197 & 0.007 & 0.211 & 0.002 & 0.149 & 0.001 & 7.0 & 0.5\\
16 Aug 10 & 2455425.685 & 0.201 & 0.004 & 0.221 & 0.002 & 0.148 & 0.003 & 4.5 & 0.5\\
02 Sep 10 & 2455442.699 & 0.189 & 0.007 & 0.218 & 0.001 & 0.147 & 0.001 & 2.6 & 0.4\\
19 Sep 10 & 2455459.473 & 0.187 & 0.005 & 0.211 & 0.002 & 0.142 & 0.001 & 1.2 & 0.4\\
26 Sep 10 & 2455466.500 & 0.198 & 0.006 & 0.215 & 0.001 & 0.147 & 0.002 & 0.1 & 0.4\\
05 Oct 10 & 2455475.518 & 0.212 & 0.007 & 0.215 & 0.002 & 0.155 & 0.002 & 1.8 & 0.7\\
13 Oct 10 & 2455483.475 & 0.230 & 0.005 & 0.213 & 0.001 & 0.165 & 0.001 & 3.6 & 0.4\\
15 Oct 10 & 2455485.832 & 0.216 & 0.003 & 0.223 & 0.002 & 0.155 & 0.002 & 3.9 & 0.4\\
17 Oct 10 & 2455486.863 & 0.222 & 0.006 & 0.217 & 0.001 & 0.154 & 0.001 & 3.8 & 0.3\\
20 Oct 10 & 2455490.471 & 0.211 & 0.016 & 0.213 & 0.001 & 0.158 & 0.001 & 5.1 & 0.8\\
12 Nov 10 & 2455513.360 & 0.199 & 0.004 & 0.212 & 0.002 & 0.149 & 0.004 & 4.9 & 0.7\\
15 Nov 10 & 2455516.831 & 0.202 & 0.003 & 0.222 & 0.002 & 0.142 & 0.003 & 4.0 & 0.4\\
20 Nov 10 & 2455521.798 & 0.197 & 0.007 & 0.212 & 0.002 & 0.137 & 0.001 & 4.2 & 0.5\\
21 Nov 10 & 2455522.718 & 0.195 & 0.006 & 0.212 & 0.004 & 0.137 & 0.004 & 3.9 & 0.4\\
26 Nov 10 & 2455527.368 & 0.185 & 0.004 & 0.212 & 0.001 & 0.136 & 0.003 & 4.5 & 0.4\\
27 Nov 10 & 2455528.855 & 0.196 & 0.006 & 0.212 & 0.002 & 0.137 & 0.002 & 3.9 & 0.4\\
04 Dec 10 & 2455535.349 & 0.175 & 0.009 & 0.207 & 0.001 & 0.138 & 0.003 & 4.8 & 0.6\\
12 dec 10 & 2455543.383 & 0.175 & 0.010 & 0.211 & 0.001 & 0.139 & 0.002 & 5.0 & 0.8\\
14 Dec 10 & 2455545.292 & 0.167 & 0.013 & 0.208 & 0.002 & 0.147 & 0.003 & 5.3 & 1.1\\
\hline                                                                
\end{tabular*}
Note: Relative intensity values are given for H$\alpha$ and CaII IR. For CaII K emission core I/I(395 nm) is measured. B$_{l}$ values and their uncertainties are given in gauss.
\end{center}
\end{table}

\begin{figure}
\begin{center}
\includegraphics[totalheight=0.33\textwidth]{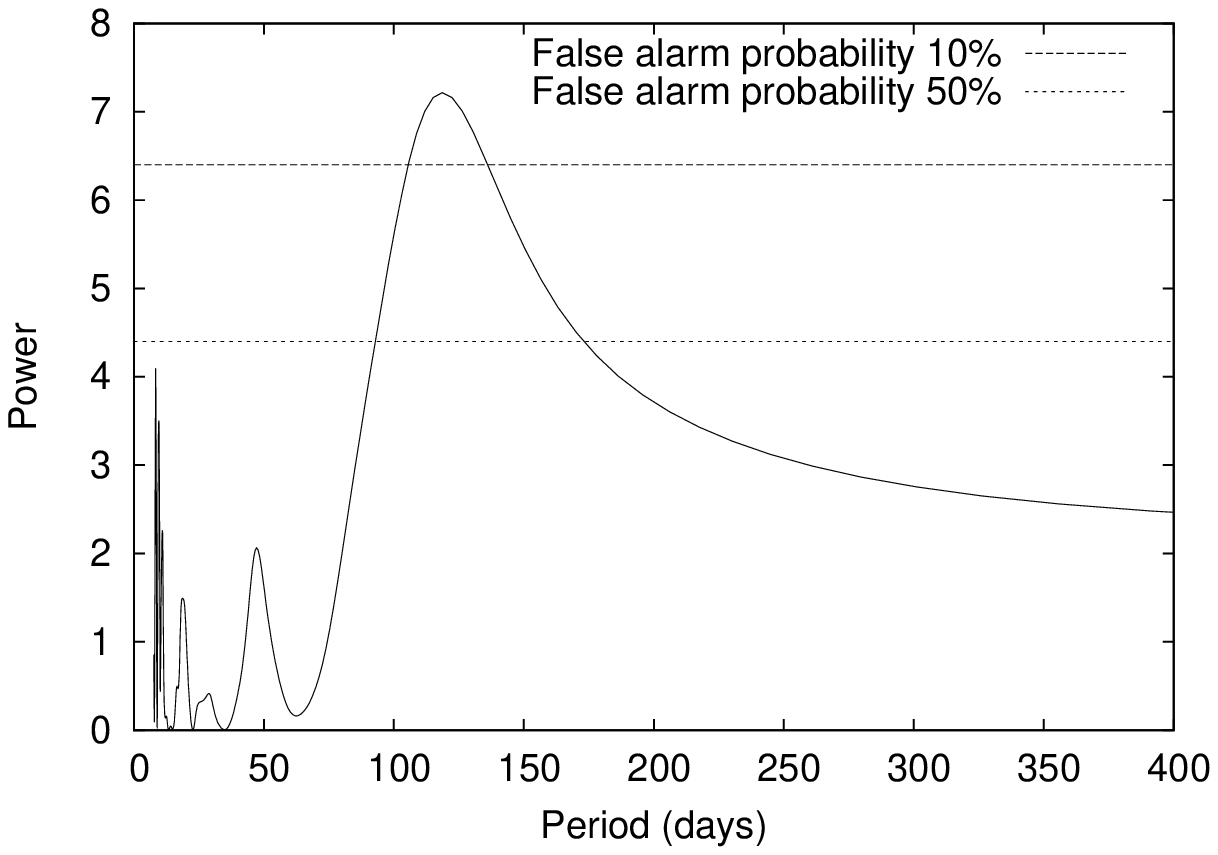}
\includegraphics[totalheight=0.33\textwidth]{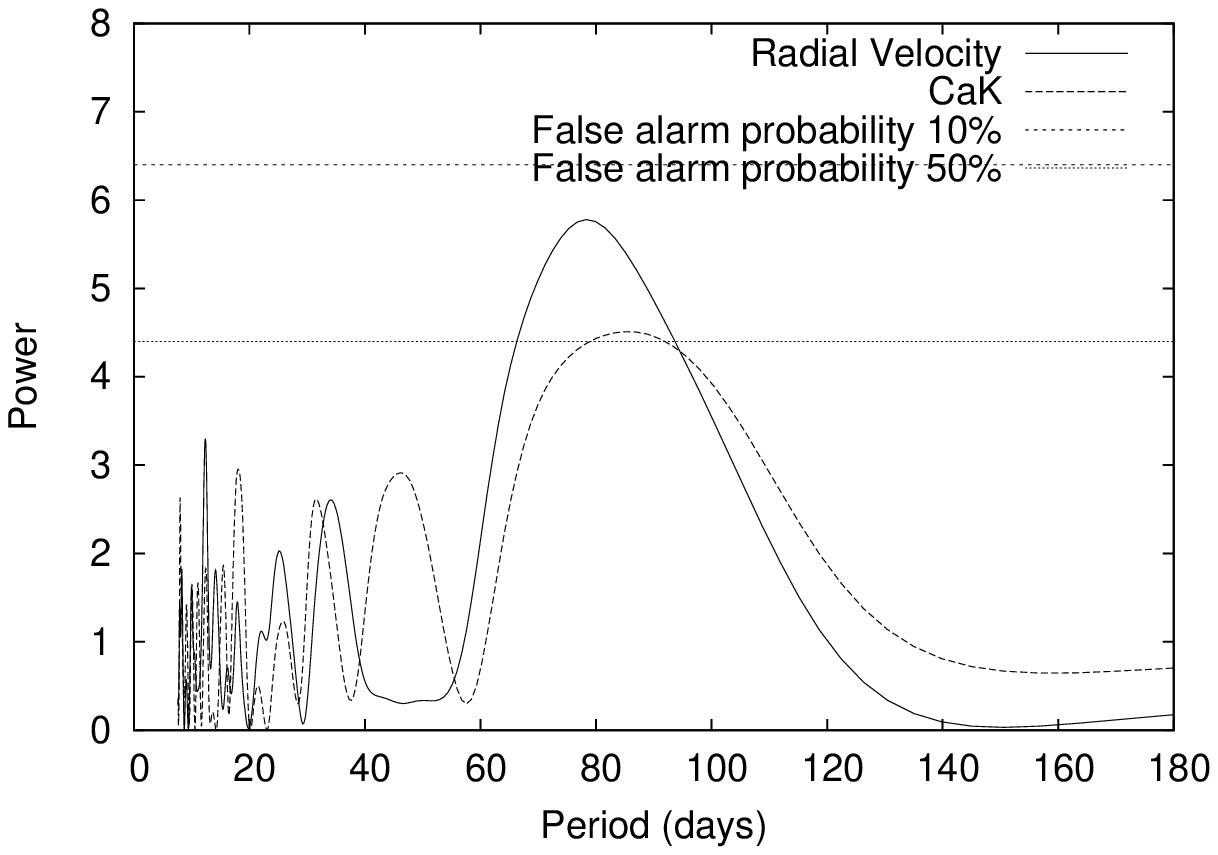}
\caption{Results from the periodogram analysis for B$_{l}$ (\textit{left panel}) and for CaII K and the radial velocity (\textit{right panel}) after applying the Lomb-Scargle method.}
\label{2figures}
\end{center}
\end{figure}

\begin{figure}
  \begin{center}
    \centering{\epsfig{file=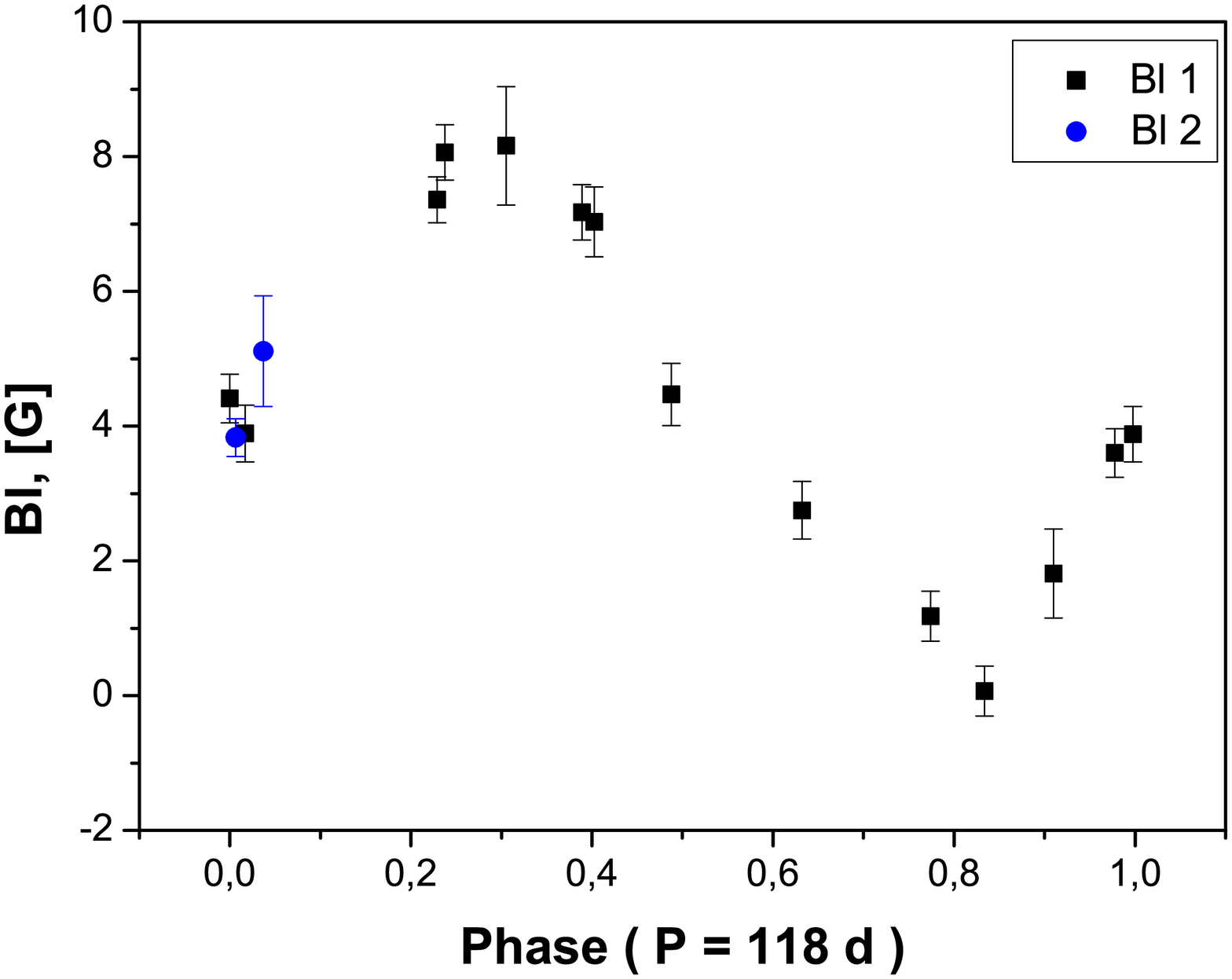, width=0.7\textwidth}}
    \caption[]{Variations of B$_{l}$ with respect to the phase with the determined possible period P = 118 days. Observations are in the period June 19, 2010 - October 20, 2010. Error bars are indicated. Black squares indicate the first rotation of the star, blue dots indicate the second rotation of the star.}
    \label{figure8}
  \end{center}
\end{figure}

\begin{figure}
  \begin{center}
    \centering{\epsfig{file=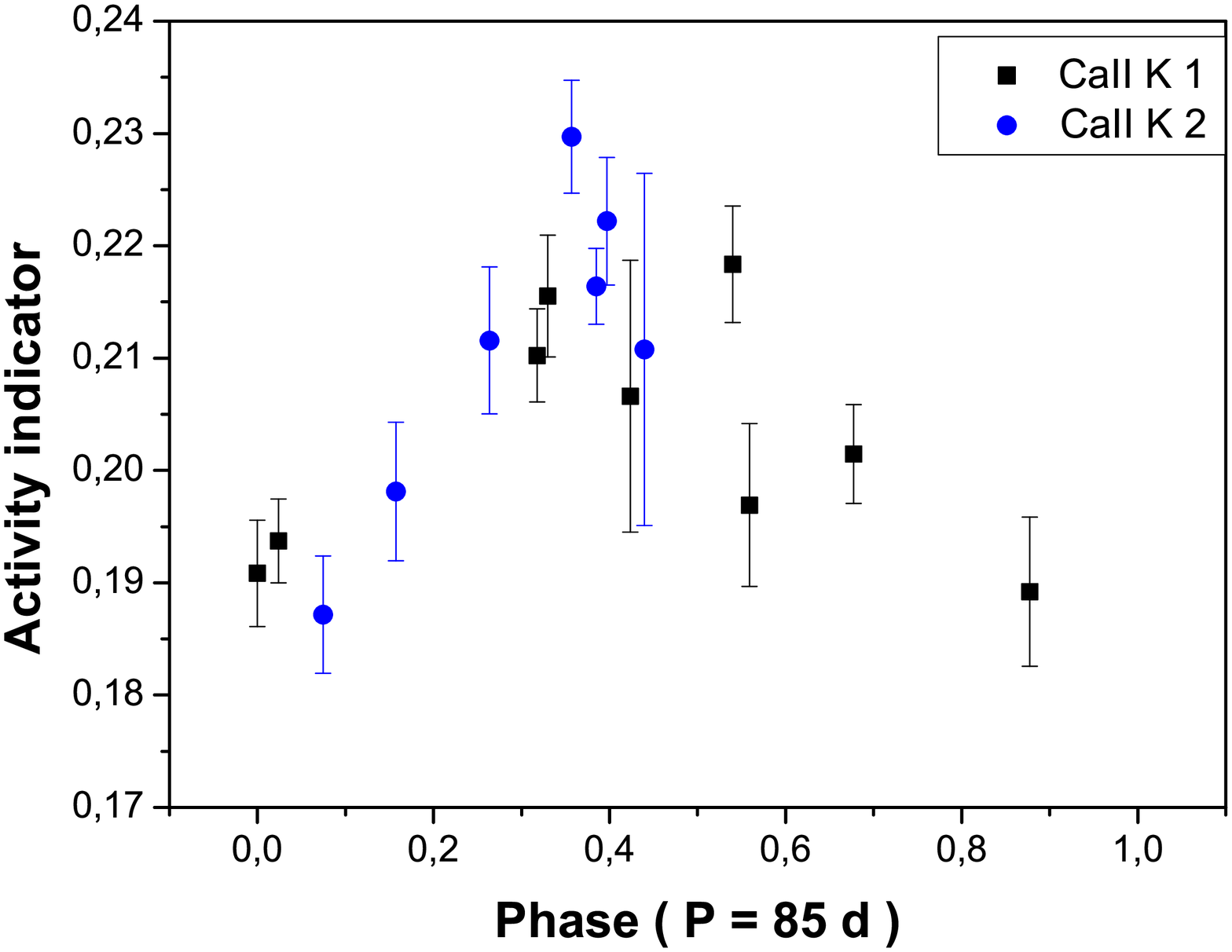, width=0.7\textwidth}}
    \caption[]{Variations of CaII K with respect to the phase with the determined possible period P = 85 days. Observations are in the period June 19, 2010 - October 20, 2010. Error bars are indicated. Black squares indicate the first rotation of the star, blue dots indicate the second rotation of the star.}
    \label{figure9}
  \end{center}
\end{figure}

\begin{figure}
  \begin{center}
    \centering{\epsfig{file=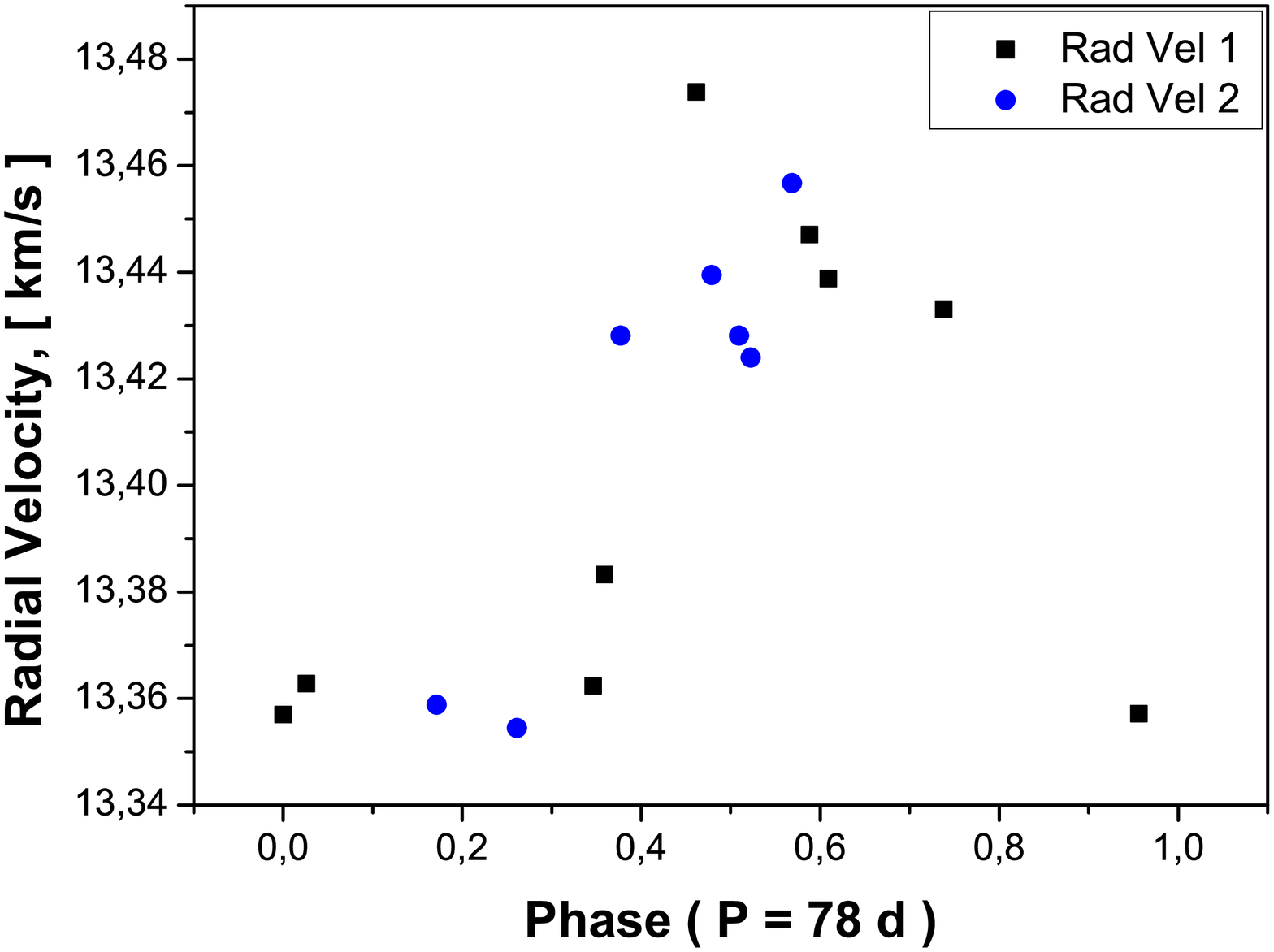, width=0.7\textwidth}}
    \caption[]{Variations of the radial velocity with respect to the phase with the determined possible period P = 78 days. The stability of the instruments is about 30 m/s (Moutou et al., 2007). Observations are in the period June 19, 2010 - October 20, 2010. Black squares indicate the first rotation of the star, blue dots indicate the second rotation of the star.}
    \label{figure10}
  \end{center}
\end{figure}

We applied a Lomb-Scargle method for the period search of the data. Because of the two different  behaviors of the dataset, it was not possible to perform the periodogram analysis with all the data. So we used only the first part of the dataset. We found possible periods for B$_{l}$, CaII K and the radial velocity.

We determined the following possible periods: for B$_{l}$ a period of P = 118 days with false alarm probability (FAP) of 5\%; for CaII K a period of P = 85 days with FAP of 51\% and for the radial velocity a period of P = 78 days with FAP of 18\%. These periods and FAP values are shown in Fig.\ref{2figures}. However, one should be cautious about these periods because our dataset covers less than two periods.

As it was mentioned, B$_{l}$ is a surface-averaged longitudinal magnetic field and it originates from the star's photosphere. The activity indicators H$\alpha$, CaII K and CaII IR originate from the star's chromosphere. Our period determinations may suggest that the chromosphere of $\beta$ Ceti rotates faster than the photosphere. Another interesting result is the RV period P = 78 days which also reflects the photospheric rotation and it is shorter than the B$_{l}$ period. We will continue to obtain data for this star in order to be able to understand better these differences.

Using these periods, we made new plots with the behavior of B$_{l}$, CaII K and RV with respect to the phase --- Fig.\ref{figure8}, Fig.\ref{figure9} and Fig.\ref{figure10}. The X--axis values for these three figures were calculated according to the determined periods respectively for B$_{l}$, CaII K and RV. These periods fit well to our observational data especially for CaII K and RV. But the data do not cover completely at least two rotations so the periods we found are possible periods. More observational data and their analysis could confirm or reject these periods.

\section*{Conclusions}
\begin{list}{-}{We collected observational data for 6 months (June 19, 2010 --- December 14, 2010) for the single late-type giant $\beta$ Ceti. The Least Squares Deconvolution method was applied for magnetic field detection and measurements. Our results are:}
\item B$_{l}$ remains of positive polarity for all our observations with variations between 0.1 G and 8.2 G.
\item We found that during the last month of our observational period a change in the configuration of the magnetic field at the stellar surface might occurs.
\item The behavior of the activity indicators follows the behavior of B$_{l}$ but only for the first part of our dataset. The case is not the same for the second part, where the emission decrease as a result that the magnetic field covers a smaller fraction of the stellar surfece.
\item Using the Lomb-Scargle method for the period search, we found possible periods for B$_{l}$ (P = 118 days), CaII K (P = 85 days) and for the radial velocity (P = 78 days).
\end{list}

\begin{list}{-}{Our next steps for studying $\beta$ Ceti are:}
\item to compute a Zeeman--Doppler Imaging map of the star
\item to collect more observational data which will help us to analyze further the magnetic variability of $\beta$ Ceti.
\end{list}

\textit{Acknowledgements} We thank the TBL and CFHT teams for providing service observing with NARVAL and ESPaDOnS. The observations in 2010 with NARVAL were funded under Bulgarian NSF grant DSAB 02/3/2010. R.K.-A., S.Ts. and R.B. acknowledge partial financial support under NSF contract DO 02-85. R.B. also acknowledges support under the RILA/EGIDE exchange program (contract RILA 05/10). G.A.W. acknowledges support from the Natural Sciences and Engineering Research Council of Canada (NSERC). S.Ts. is thankful for the possibility to work for 3 months in 2011 in IRAP, Tarbes, France, under the Erasmus program.


\end{document}